\documentclass[journal=cdgefu,manuscript=article]{achemso}
\usepackage[version=3]{mhchem}
\usepackage{graphicx}
\usepackage{bm}
\usepackage{amsmath}
\usepackage{amssymb}

\author{Rafaela F. S. Penacchio}
\affiliation{Institute of Physics, University of S{\~{a}}o Paulo, 05508-090 S{\~{a}}o Paulo, SP, Brazil}

\author{Celso I. Fornari}
\affiliation{Experimentelle Physik VII and W{\"u}rzburg Dresden Cluster of Excellence ct.qmat, Fakult{\"a}t f{\"u}r Physik und Astronomie, Universit{\"a}t W{\"u}rzburg, Am Hubland, D-97074 W{\"u}rzburg, Germany}

\author{Yor{\'i} G. Camillo}
\affiliation{Institute of Physics, University of S{\~{a}}o Paulo, 05508-090 S{\~{a}}o Paulo, SP, Brazil}

\author{Philipp Kagerer}
\affiliation{Experimentelle Physik VII and W{\"u}rzburg Dresden Cluster of Excellence ct.qmat, Fakult{\"a}t f{\"u}r Physik und Astronomie, Universit{\"a}t W{\"u}rzburg, Am Hubland, D-97074 W{\"u}rzburg, Germany}

\author{Sebastian Buchberger}
\affiliation{Experimentelle Physik VII and W{\"u}rzburg Dresden Cluster of Excellence ct.qmat, Fakult{\"a}t f{\"u}r Physik und Astronomie, Universit{\"a}t W{\"u}rzburg, Am Hubland, D-97074 W{\"u}rzburg, Germany}

\author{Martin Kamp}
\affiliation{Physikalisches Institut and R{\"{o}}ntgen-Center for Complex Material Systems (RCCM), Fakult{\"{a}}tf{\"{u}}r Physik und Astronomie, Universit{\"{a}}tW{\"{u}}rzburg, W{\"{u}}rzburg D-97074, Germany}

\author{Hendrik Bentmann}
\affiliation{Experimentelle Physik VII and W{\"u}rzburg Dresden Cluster of Excellence ct.qmat, Fakult{\"a}t f{\"u}r Physik und Astronomie, Universit{\"a}t W{\"u}rzburg, Am Hubland, D-97074 W{\"u}rzburg, Germany}

\author{Friedrich Reinert}
\affiliation{Experimentelle Physik VII and W{\"u}rzburg Dresden Cluster of Excellence ct.qmat, Fakult{\"a}t f{\"u}r Physik und Astronomie, Universit{\"a}t W{\"u}rzburg, Am Hubland, D-97074 W{\"u}rzburg, Germany}

\author{S{\'{e}}rgio L. Morelh{\~{a}}o}
\affiliation{Institute of Physics, University of S{\~{a}}o Paulo, 05508-090 S{\~{a}}o Paulo, SP, Brazil}
\email{morelhao@if.usp.br}

\title{Simulation of X-ray diffraction in \ce{Mn$_x$Bi2Te$_{3+x}$} epitaxic films}

\begin{document}

\begin{abstract}
Disordered heterostructures stand as a general description for compounds that are part of homologous series such as bismuth chalcogenides. In device engineering, van der Waals epitaxy of these compounds is very promising for applications in spintronic and quantum computing. Structural analysis methods are essential to control and improve their synthesis in the form of thin films. Recently, X-rays tools have been proposed for structural modeling of disordered heterostructures [arxiv.org/abs/2107.12280]. Here, we further evaluate the use of these tools to study the compound \ce{Mn$_x$Bi2Te$_{3+x}$} in the grazing incidence region of the reflectivity curves, as well as the effect of thickness fluctuation in the wide angle region.  
\end{abstract}

\section{Grazing Incidence X-ray Reflectometry}\label{graz}

\begin{figure*}
    \centering
    \includegraphics[width=.75\textwidth]{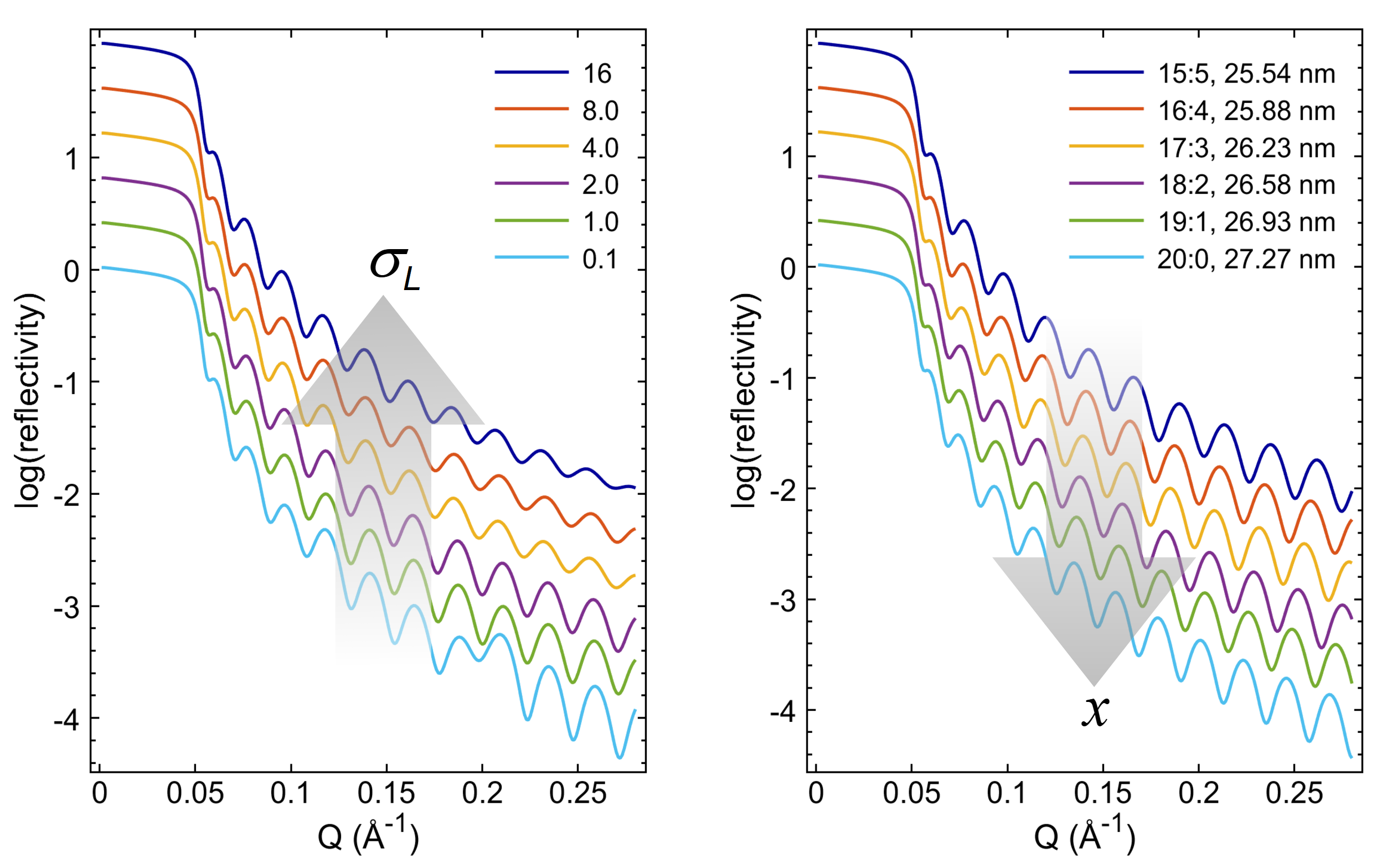}
    \caption{X-ray reflectivity simulation in (\ce{MnBi2Te4})$_n$(\ce{Bi2Te3})$_m$ heterostructures on \ce{BaF2}(001) at grazing incidence angles, $\theta\!<\!5^\circ$ for Cu$K_\alpha$ radiation (wavelength $\lambda = 1.5418\textrm{\AA}^{-1}$). $Q = (4\pi/\lambda)\sin\theta$. Left panel, $16$:$4$ heterostructures with disorder $\sigma_L$, in eq.~(2) of Ref.~\citenum{rp21}, varying from 0.1 to 16 as indicated. Right panel, $n$:$m$ heterostructures as a function of composition $x = n/(n+m)$ and constant disorder $\sigma_L=1$. Total film thickness $T = n\, d_{SL} + m\, d_{QL}$ is also indicated where $d_{SL}=1.364$\,nm and $d_{QL}=1.0165$\,nm (see Table I of Ref.~\citenum{rp21}).}
    \label{fig:AnBmGz}
\end{figure*}

Diffraction of X-rays in layered materials of large $d$-spacing can be exactly calculated by means of a set of recursive equations described in details elsewhere \cite{sm17a,rp21}. It is valid in grazing incidence specular reflection geometries \cite{sm02b,sm02c}, as well as near strong Bragg reflections from the substrate lattice undergoind dynamical diffraction. In Fig.~\ref{fig:AnBmGz}, it is demonstrated for (\ce{MnBi2Te4})$_n$(\ce{Bi2Te3})$_m$ heterostructures as a function of the degree of disorder and composition $x = n/(n+m)$ of the \ce{Mn$_x$Bi2Te$_{3+x}$} (MBT) film. Using ensemble of model structures of disordered  heterostructures, as given in Ref.~\citenum{rp21}, is a method quite limited for X-ray reflectivity simulation at grazing incidence angles. In actual films, this method of adding diffracted intensities from many statistically equivalent models is accurate for films with large crystallographic domains where interference effects between laterally adjacent domains can be neglected. The method become more reliable at wide angles where these interference effects are minimized. A more detailed discussion on this subject can be found elsewhere \cite{sm17a}. This set of recursive equations that become very suitable for layered materials has emerged within a vast effort to advance X-ray diffraction methods for analyzing relevant materials, ranging from nanostructured devices to biological tissues \cite{sm91,la98,sm02a,sm03,aa06,am10,sm11,sm18}.

\section{Thickness Fluctuation}\label{thick}

The quality of epitaxial films is strongly dictated by lattice matching. In van der Waals (vdW) epitaxy where week interlayer forces are responsible for the film structure along the growth direction, films of good quality can be achieved even in cases where there are misfits of a few percent \cite{ag13,pv18,jh17,gs18,mi19}. However, the misfit between film and substrate in-plane lattice parameters impacts the lateral lattice coherence length, that is the size of perfect crystallographic domains \cite{sm19}. Small uncorrelated domains lead to thickness fluctuation that can be inferred by the absence of thickness fringes in the specular reflectivity curves. At grazing angles there is the problem that severe thickness inhomogeneity can compromise data analysis. Fortunately, X-ray diffraction simulation of long $Q$ scans in (\ce{MnBi2Te4})$_n$(\ce{Bi2Te3})$_m$ heterostructures have revealed a few reflections, such as the $0\,0\,0\,24$ reflection ($L24$), around which thickness fringes are visible regardless disorder and composition \cite{rp21}.    

\begin{figure}
    \centering
    \includegraphics[width=0.75\textwidth]{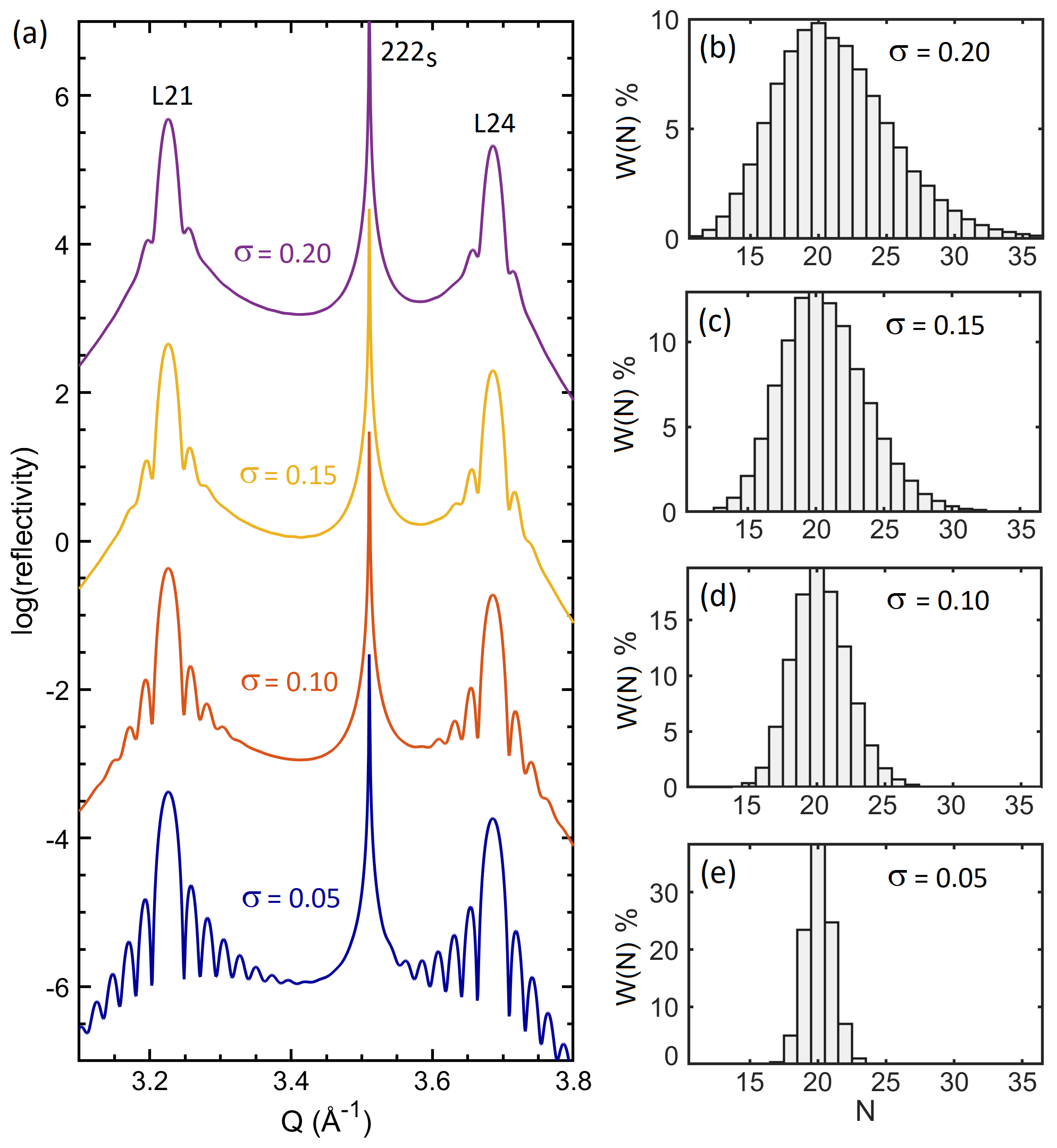}
    \caption{(a) X-ray reflectivity simulation in \ce{MnBi2Te4} films on \ce{BaF2} (111) substrate as a function of thickness distribution over the film area. $L21$ and $L24$ stand for film reflections that appear around the substrate 222 reflection. $Q = (4\pi/\lambda)\sin\theta$. (b-e) Weight function $W(N)$ for intensity contribution of film regions with $N$ SLs. The most probable number of SLs in the films is $N_0=20$ and the value $\sigma$ in Eq.~(\ref{eq:Lx}) are indicated in each case, as well as near the corresponding reflectivity curve. }
\label{fig:thick}
\end{figure}

To estimate the amount of thickness fluctuation necessary for eliminating the fringes around the $L24$ peak, a weight function 
\begin{equation}\label{eq:WofN}
    W(N) = \int_N^{N+1} L(x) dx\,,
\end{equation}
is used to account for the intensity contribution of film areas containing $N$ \ce{MnBi2Te4} septuble layers (SLs). 
\begin{equation}\label{eq:Lx}
   L(x) = \frac{1}{x\, \sigma \sqrt{2\pi}} {\rm exp}\left \{-\frac{[{\rm ln}(x)-{\rm ln}(b) ]^2}{2\sigma^2}\right \}
\end{equation}
where $b = N_0 \exp(\sigma^2)$, $N_0$ is the most probable thickness (mode), and $\sigma$ is the standard deviation in log scale. The simulated reflectivity curves in a $Q$ range near the $L24$ peak are shown in Fig.~\ref{fig:thick}(a) and the corresponding thickness distributions in Figs.~\ref{fig:thick}(b-e). For films with most probable thickness around 20 SLs, the fringes vanish when the weights above 1\% are distributed from $N=13$ to 31 SLs.  

\section{Reflection and Transmission Coefficients}\label{RTCoeff}

An X-ray monochromatic plane-wave as it crosses an atomic plane undergoes reflection, refraction, and absorption. In standard kinematical theory of X-ray diffraction, only the first of these phenomena is accounted for. On the other hand, the dynamical diffraction theory besides include these three phenomena, also takes into account the rescattering process between successive planes of atoms. In a more general approach, the dynamical diffraction theory can embrace more complex situations where many reflections are fully excited simultaneously, within the so-called multi-wave diffraction configuration \cite{ew97,sm17b}. However, for investigating Bragg reflections under specular diffraction geometry in layered materials, the theory can be significantly simplified in terms of the recursive series \cite{rp21} where the reflected and transmitted amplitudes of the X-ray waves by atomic monolayers (MLs) are given in terms of the reflection and transmission coefficients,
\begin{equation}\label{eq:rxtx}
r_X = -i\frac{r_e \lambda C}{\sin\theta}\sum_a\eta_a f_a(Q,E)\quad{\rm and}\quad t_X = 1+i\frac{r_e \lambda}{\sin\theta}\sum_a\eta_a f_a(0,E)\,,   
\end{equation}
respectively. $Q = (4\pi/\lambda) \sin \theta$ is the modulus of the diffraction vector along the MLs normal direction given as a function of the incidence angle $\theta$, scattering angle $2\theta$, and wavelength $\lambda$. $\eta_a$ is the area density of atoms $a$ in the ML plane and $f_a(Q,E)=f_a^0(Q)+f_a^{\prime}(E)+if_a^{\prime\prime}(E)$ are their atomic scattering factors with resonant amplitudes $f_a^{\prime}(E)$ and $f_a^{\prime\prime}(E)$ that are a function only of the X-ray photon energy $E$. $r_e = 2.818\times10^{-5}$\,\AA\, is the classical electron radius from the Thomson elastic scattering and photoelectric absorption cross sections \cite{ja11}. The $\sin\theta$ in the denominator of Eq.~(\ref{eq:rxtx}) takes into account area variation of the beam footprint at the sample surface, and the polarization term $C=\sqrt{(1+\cos^2 2\theta)/2\,}$ stands for unpolarized X-rays. Due to the collimating optics of the used diffractometer as well as the $\sigma$-polarization available in most synchrotron diffraction stations, $C=1$ has been considered throughout this work. 

\begin{figure}
    \centering
    \includegraphics[width = .5\textwidth]{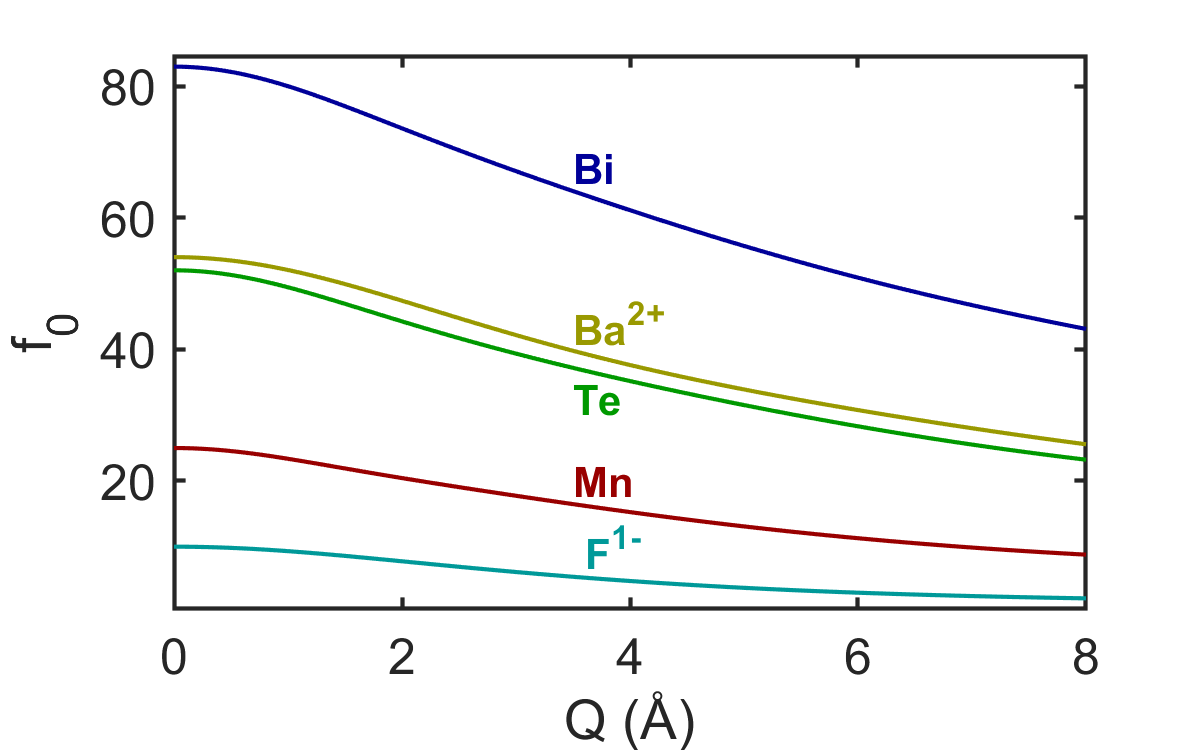}
    \caption{Angle dependent component of atomic scattering amplitudes for x-rays $f_0(Q)$, given in number of electrons. Exact values were obtained from the \texttt{asfQ.m} codes \cite{sm16}.}
    \label{fig:asfQ}
\end{figure}

From the perspective of the 00L film reflections and $hhh$ substrate reflections, each ML contains only one type of chemical element, that is the MLs are monoatomic layers. Basically there are the Mn, Bi, and Te MLs in the film and the Ba$^{2+}$ and F$^{1-}$ MLs in the substrate. The actual values of non-resonant atomic scattering amplitudes $f_a^0(Q)$, are shown in Fig.~\ref{fig:asfQ} as a function only of $Q$ within the spherosymmetric atom approximation \cite{sm16}. No ionic charges were considered for the metallic atoms. Table~\ref{tab:fpfpp} provides the resonant amplitudes values used in the X-ray diffraction simulations. For the area density of atoms in the MLs, $\eta_a$ in Eq.~(\ref{eq:rxtx}), the used values where $\eta_{\rm Ba} = \eta_{\rm F} = 2/a_{\rm S} \sin(60^\circ) = 6.0076\times10^{-2}\,\textrm{atoms/\AA}^2$ in the substrate MLs, $\eta_{\rm Bi} = \eta_{\rm Te} = 1/a_{\rm QL} \sin(120^\circ) = 6.0025\times10^{-2}\,\textrm{atoms/\AA}^2$ in the \ce{Bi2Te3} quintuple layers (QLs), and 
$\eta_{\rm Mn} = \eta_{\rm Bi} = \eta_{\rm Te} = 1/a_{\rm SL} \sin(120^\circ) = 6.1483\times10^{-2}\,\textrm{atoms/\AA}^2$ in the \ce{MnBi2Te4} septuple layers (SLs).

\begin{table}[h]
    \centering
    \begin{tabular}{rccccc}
    \hline\hline
    atom: & Mn & Bi & Te & Ba & F \\
    \hline
    $f_a^{\prime} =$ & $-0.532920$ & $-3.877040$ & $-0.153572$ & $-1.015625$ & $+0.073181$ \\
    $f_a^{\prime\prime}=$ & $+2.808637$ & $+8.937114$ & $+6.358851$ & $+8.469466$ & $+0.053438$ \\
        \hline\hline
\end{tabular}
    \caption{Resonant atomic scattering amplitues obtained from the \texttt{fpfpp.m} codes \cite{sm16} for the used energy $E=8041.57$\,eV ($\lambda = 1.5418$\,\AA).}
    \label{tab:fpfpp}
\end{table}

\bibliography{arXiv2021XRDsimulation}

\end{document}